\documentclass[12pt]{article}
\usepackage{graphicx}
\usepackage{bm}
\usepackage[width=\linewidth]{caption}
\usepackage{subcaption}
\usepackage{amsmath}
\usepackage{xcolor}
\usepackage{array}
\usepackage{vcell}
\usepackage{tikz}
\usepackage{url}
\usepackage[backend=biber]{biblatex}
\usepackage[letterpaper, margin=1in]{geometry}

\newcommand{\VarTime}{t}

\newcommand{\bfx}{\bm{x}}

\newcommand{\refsec}[1]{Section \ref{#1}}
\newcommand{\reffig}[1]{Figure \ref{#1}}
\newcommand{\reftable}[1]{Table \ref{#1}}

\title{Optimal design of triply-periodic minimal surface implants for bone repair}
\author{David Cohen \and Juli\'an A. Norato\footnote{Corresponding author: julian.norato@uconn.edu}}

\date{School of Mechanical, Aerospace, and Manufacturing Engineering \\ University of Connecticut, Storrs, CT USA 06269}

\begin{document}

\maketitle

\abstract
This work proposes a gradient-based method to design bone implants using triply-periodic minimal surfaces (TPMS) of spatially varying thickness to maximize bone in-growth. Bone growth into the implant is estimated using a finite element based mechanobiological model considering the magnitude and frequency of in vivo loads, as well as the density distribution of the surrounding bone. The wall thicknesses of the implant unit cells are determined via linear interpolation of the thicknesses over a user defined grid of control points, avoiding mesh dependency and providing control over the sensitivity computation costs. The TPMS structure is modeled as a homogenized material to reduce computational cost. Local properties of the implant are determined at run-time on an element-by-element basis using a pre-constructed surrogate model of the TPMS’s physical and geometric properties as a function of the local wall thickness and the density of in-grown bone. Design sensitivities of the bone growth within the implant are computed using the direct sensitivity method. The methodology is demonstrated on a cementless hip, optimizing the implant for bone growth subject to wall thickness constraints to ensure manufacturability and allow cell infiltration.
\endabstract

\section{Acknowledgments}
This work was supported by the National Science Foundation, award CMMI-1727591.

\section{Statements and Declarations}
The authors have no competing interests to declare.

\section{Introduction}
Hip arthroplasties, colloquially known as hip replacements, are a common surgical procedure where all or part of a patient's hip joint is replaced with a prosthesis. Over 250,000 of these procedures were performed in the United States alone in 2019, with more than 500,000 a year expected by 2040 \cite{Shichman2023a}. While the majority of hip replacements require no further surgery after the implantation, revision surgery is unfortunately required in tens of thousands of cases each year. These revision surgeries entail considerable cost, both monetary and in terms of time to recuperate. Moreover, they carry a greater risk of complications than primary  hip arthroplasties \cite{Patel2023, Guo2022}. One of the leading causes of revision is the lack of adequate implant fixation, resulting in aseptic loosening or instability of the implant. Studies suggest these these two failure modes account for more than $30\%$ of all revisions \cite{Ulrich2008, Hinton2022, Patel2023, Upfill-Brown2021}. Improving implant fixation thus presents a promising avenue for reducing revision surgeries.

There are two main methods of securing the hip prosthesis to the bone. The first is to use a special cement, typically a form of acrylic, to bond the implant to the bone. The cement fills the gap between the implant and bone, seeping into pores in the latter to physically interlock the pair once cured \cite{Maggs2018}. The second method instead relies on the bone itself to provide fixation. The implant is initially press fit into place and its surface is designed to allow the existing bone to grow into it, locking it in place. In the US, these cementless implants are dominant, accounting 95\% of all hip arthroplasties in 2023 \cite{Ryan2024}. This work focuses specifically on this type.

Implants may come loose due the absence of growth or outright bone loss after surgery. One factor that contributes to this is the lack of adequate mechanical stimulus from load, which is needed to grow and maintain bone. Hip implants are typically made of materials far stiffer than bone and they are often solid, which means they will take a significant portion of the load and consequently reduce the loading and hence the mechanical stimulus on the surrounding bone. This phenomenon is referred to as stress shielding \cite{Fraldi2010}, and causes bone to begin to atrophy. To avoid stress shielding, a common research goal for implant design techniques is thus to mimic the stress distribution of the intact bone in the remaining bone. Another cause of implant loosening is slippage at the implant-bone interface. While some slippage is inevitable due to deformation of the bone and implant under cyclic loading, micromotions of greater 40 $\mu$m seem to reduce bone growth, and those than 150 $\mu$m completely inhibit it \cite{Pettersen2009}. A design strategy to circumvent slippage is to promote osteointegration, which minimizes relative motion between the bone and the implant.

To improve the clinical outcomes of hip implants, a number of authors have turned to computational optimization techniques to refine their designs. Many of these works are based at least partially on conventional density based topology optimization \cite{Bendsøe2004}, and seek to minimize the implant compliance subject to a volume fraction constraint \cite{Fraldi2010, Tan2021, He2018, Moussa2020}. In \cite{Fraldi2010}, the stress field in the femur with the optimized implant was compared to those in the intact femur for the same loads, and with those of an unoptimized reference design. The optimized designs showed a smaller difference in the femoral stresses relative to the stresses in the intact bone, suggesting reduced stress shielding. In \cite{Tan2021}, a reference implant is optimized with and without density filters. In the latter case, the intermediate densities are interpreted as material with varying porosity. The designs are then evaluated via finite element analysis (FEA) to determine the interface strains under in-vivo loads, which are subsequently compared to those of the reference implant and intact femur. The more compliant optimized designs were found to have strains more similar to those of the intact bone than the reference. Similar to \cite{Tan2021}, \cite{Moussa2020} uses an unfiltered density-based optimization to create a functionally-graded structure consisting of periodic unit cells, which are then evaluated in terms of maximum contract stress and interface strain compared to reference designs that are solid and uniformly porous. The optimized design has lower stresses and strains than the solid design, but is roughly equivalent to the uniformly porous design. The authors of \cite{He2018} uses density-based topology optimization as a first step. Additional constraints are placed on the regional compliance in several regions of the bone to control stress shielding in the first stage of the proposed method. Then, this design is used as a template for a lattice optimization. High-density regions are interpreted as solid material, whereas low-density regions correspond to a lattice of periodic unit cells. The diameter of the lattice struts is then optimized to minimize material usage subject to stress constraints. The final optimized design was found to better replicate the stress field in the intact femur.  

In addition to topology optimization, a number of works using shape optimization for femur implants have also been published \cite{Nicolella2006, Ruben2007, Ruben2012}. The first is a probabilistic shape optimization that seeks to minimize the overall probability of failure of the implant given the uncertainties in the material properties and loads of the femur-implant system. The optimized design reduced the probability of the various modes of failure investigated by 30-99\%. The latter two papers are multi-objective optimizations that minimize the implant contact stress, tangential displacements at the interface \cite{Ruben2007}, and maximize the total strain energy in the bone \cite{Ruben2012}. The optimization was performed several times with different objective weights, correspondingly yielding different designs. Interestingly, \cite{Ruben2012} also assessed the long term implant fixation using a simple bone growth model to see if bone would grow into the implant surface. The optimized designs had mixed results compared to the reference design for bone growth, which was neither a constraint nor objective in the optimizations. This is one of the only works which considers bone growth into the implant.

In this work, we propose a novel method to optimize bone implants made of a triply-periodic minimal surface (TPMS) lattice to maximize bone ingrowth. Sheet-network TPMS lattices are here considered due to their high internal surface area-to-volume ratio, which increases the available area for bone to be deposited on; their interconnected void region, which facilitates bone ingrowth; and their minimal-surface property, which precludes the need for internal supports if fabricated via additive manufacturing.  The shape of the implant is considered fixed, and the optimization determines the optimal distribution of the lattice thickness within the implant. A bone growth model previously presented by the authors adapted from the model of \cite{Sanz-Herrera2008} is used to estimate bone growth into the femur based on the frequency and force of in-vivo loads, the bone geometry and density distribution, and the geometry of the implant \cite{Cohen2021, Cohen2022}.  Analytical design sensitivities of the bone growth are computed to enable the use of efficient gradient-based optimizers. To demonstrate the proposed technique, we optimize the femoral component of a titanium cementless hip implant made of a gyroid TPMS lattice structure for the stem. 

The bone growth model and its corresponding design sensitivities, and the formulation of the optimization problem are detailed in \refsec{sec:formulation}. The setup of the finite element analysis problem is discussed in \refsec{sec:implementation}. \refsec{sec:results} presents results of the optimization, and we draw conclusions of our work in \refsec{sec:conclusion}.

\section{Formulation} \label{sec:formulation}

\subsection{Bone growth model} \label{subsec:bone-growth-model}

The bone growth model used in this work is based on the model of \cite{Sanz-Herrera2008}, and it was adapted by the present authors to ensure differentiability of the bone growth function and therefore allow the use of efficient gradient-based methods for the optimization \cite{Cohen2021, Cohen2022, Cohen2024}. The growth model adopted here is unchanged from these prior works, and we include a high-level description here for completeness. The interested reader is referred to \cite{Cohen2024} for a detailed presentation of the model.

The creation of new bone is performed by specialized cells called osteoblasts, which deposit a matrix composed of mostly collagen, calcium phosphate and calcium carbonate in response to mechanical stimulus in their substrate. This is a metabolically demanding process, which requires an adequate blood supply to deliver the necessary oxygen and nutrients to the growth front \cite{Carter2001}. Thus bone growth requires the satisfaction of the following \textit{conditions}: 1) a surface on which bone deposition can occur, 2) osteoblasts on the surface to produce the new bone, 3) vascularization to supply the osteoblasts, and 4) mechanical deformation of the surface to stimulate the osteoblasts into bone deposition. 

The bone growth model estimates the change in the bone density by performing a linear static finite element analysis of the bone-implant system under typical in-vivo loads. The model quantities are computed at each element centroid and are assumed to be uniform within the element. The elemental strain energy density and frequency of each load case are used to compute the mechanical stimulus. This stimulus is subsequently combined with the amount of surface area available for deposition at the element relative to its volume (i.e., its specific surface area) to determine how much bone grows in the element. Therefore, the model considers how well each element satisfies the aforementioned surface and stimulus conditions (i.e., conds.~1 and 4, respectively). The vascularization condition (i.e., cond.~3) is addressed by the choice of type of lattice and by the imposition of geometric constraints, as outlined in \refsec{subsec:gyroid-properties}. Condition 2 (osteoblasts availability) is assumed satisfied everywhere in the bone-implant system within the first time-step after implantation. The deposition of new tissue alters both the elastic properties and specific surface area of each element, which results in the rate of bone growth changing over time. Consequently, a transient analysis is performed whereby the bone-implant system properties are updated and the bone growth analysis performed over multiple time steps to determine the accumulated bone growth in the system over a specified simulation period. The rate and thresholds for bone deposition are tuned using species-dependent empirically determined constants \cite{Carter2001, Sanz-Herrera2008}.

The analysis region $\Omega$ is separated into the mutually exclusive implant ($\Omega_d$), inert ($\Omega_{i}$), and bone ($\Omega_b$) domains. The inert domain, $\Omega_{i}$, contains regions where no bone growth is assumed to occur, such as those composed of solid metal. As its name suggests, the bone domain encompasses all of the bone not within the implant. Conversely, the implant itself and any bone that grows within it fall in the implant domain. Any elements in the bone and implant domains are henceforth referred to as bone and implant elements, respectively. The calculations pertaining to these domains are detailed in \cite{Cohen2021} and are not repeated here for brevity. For completeness, however, the following description outlines the basics of the density update procedure for both regions.

The density update at each element $e$'s centroid $\bm{x} \in \Omega$ is given by
\begin{equation}
    \label{eq:density-update}
    {\rho}_{\VarTime+\Delta \VarTime}(\bfx) = \widehat{\min}(\hat{\rho}_b, \, {\rho}_{\VarTime}(\bfx) +\dot{\rho}(\bfx)\Delta \VarTime),
\end{equation}
where ${\rho}_{\VarTime+\Delta \VarTime}$ and ${\rho}_\VarTime$ denote the bone density at times $\VarTime+\Delta \VarTime$ and $t$, respectively, and $\hat{\rho}_b$ is the maximum possible density of bone. This density is based on empirical measurements of bone density. The function $ \widehat{\min}$ is a continuously differentiable approximation of the exact minimum. The implant is assumed to have no bone inside it at implantation, so we set $\rho_0 = \tilde{\rho}_b$ for all $\bfx \in \Omega_s$.  To prevent an ill-posed mechanical analysis and preclude divisions by zero in the growth model (see \eqref{eq:mech-stimulus}), a small minimum allowable bone density $\tilde{\rho}_b$ is enforced everywhere. Numeric values for both  $\tilde{\rho}_b$ and $\hat{\rho}_b$ can be found in \reftable{tab:growth-constants}.

At every time step, the rate of bone density change $\dot{\rho}(\bfx)$ is computed as \cite{Sanz-Herrera2008}
\begin{equation}
\label{eq:dot-rho}
\dot{\rho}(\bfx)=S_{eff} S(\bfx) \dot{r}(\bfx)  \hat{\rho}_{b},
\end{equation}
where $S(\bfx)$ is the specific surface area in element $e$. 
$S_{eff}$ represents the fraction of exposed surface area where bone deposition occurs, and it is an empirical constant. Both $S(\bfx)$ and $S_{eff}$ are different in the bone and implant domains. In the bone domain, the specific surface area of bone is determined using an experimentally derived relationship $S_{b}(\bfx)$ detailed in \cite{Jacobs1994, Cohen2021}. Thus for $S(\bfx \in \Omega_b) = S_{b}(\bfx)$. In the implant domain, it is

\begin{equation}
\label{eq:ssa-implant}
S(\bfx \in \Omega_d) = \frac{\tilde{\rho}_{b}}{\rho_{\VarTime}(\bfx)}S_{d}( \bfx)+S_{b}(\bfx),
\end{equation}

where $S_{d}(\bfx)$ is the specific surface area of the empty implant with no internal bone growth. $S_{d}$ is determined at via surrogate, discussed further in \refsec{subsec:gyroid-properties}. $\dot{r}(\bfx)$ is the rate of bone deposition, calculated based on the mechanical stimulus at point $\bfx$. It is assumed bone resorption is negligible over the growth period, hence $\dot{\rho}(\bfx) \geq 0$.

The bone density growth rate is calculated as
\newcommand{\Heaviside}{\hat{H}}
\begin{equation} \label{eq:dot-r}
    \dot{r}(\bfx) = c_{s}\Delta \Psi(\bfx) \Heaviside(\Delta \Psi(\bfx)),
\end{equation}
where 
\begin{equation}
    \Delta \Psi(\bfx) = \Psi(\bfx) - \Psi_{*} - w.
\end{equation}
where $\dot{r}$ is the deposition rate of bone in units
of length per unit time; $\Psi$ is the mechanical
stimulus at element $e$; $\Psi^{*}$ is the reference stimulus level;
 $w$ half-width of the dead band around $\Psi^{*}$ where no change in bone density occurs; and $c_{s}$ is a empirical constant to tune the growth rate. All these constants are provided in \reftable{tab:growth-constants}. $\Heaviside$ is a differentiable Heaviside approximation (cf ~\cite{Cohen2024}).

 The mechanical stimulus for $N$ loadings is found via \cite{Carter2001}:
\begin{equation}
\label{eq:mech-stimulus}
\Psi(\bfx)=\left(\frac{\hat{\rho}_{b}}{\rho_{\VarTime}(\bfx)}\right)^{2}\left(\sum_{i=1}^{N}n_{i}\bar{\sigma}_{i}(\bfx)^{m}\right)^{\frac{1}{m}},
\end{equation}
where $n_{i}$ is the number of cycles of loading $i$ the bone experiences per unit time, $\bar{\sigma}_{i}$
is the effective stress for loading $i$,  and $m$ is an empirical weighting factor \cite{Carter2001}. Note that this measure of stimulus is similar to a $p$-norm smooth maximum, and thus is dominated by the largest magnitude and most frequent load.

\subsection{Lattice properties} \label{subsec:gyroid-properties}

\begin{figure} 
    \centering
    \includegraphics[width=0.5\linewidth]{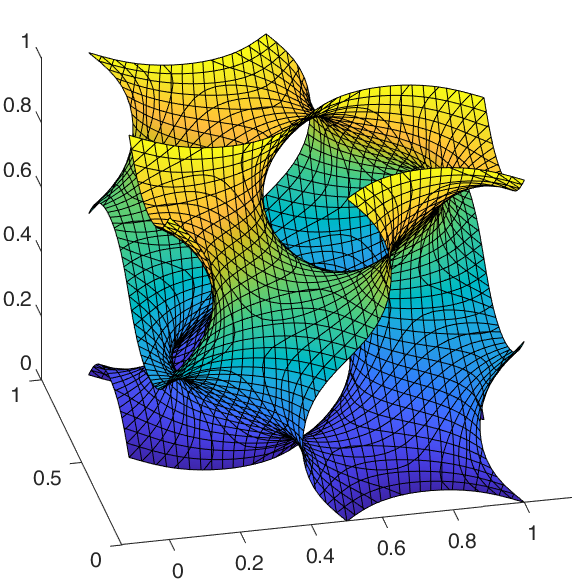}
    \caption{Gyroid surface unit cell}
    \label{fig:gyroid-surface}
\end{figure}

\begin{figure} 
    \centering
    \begin{subfigure}[b]{0.25\linewidth}
        \includegraphics[width=\linewidth]{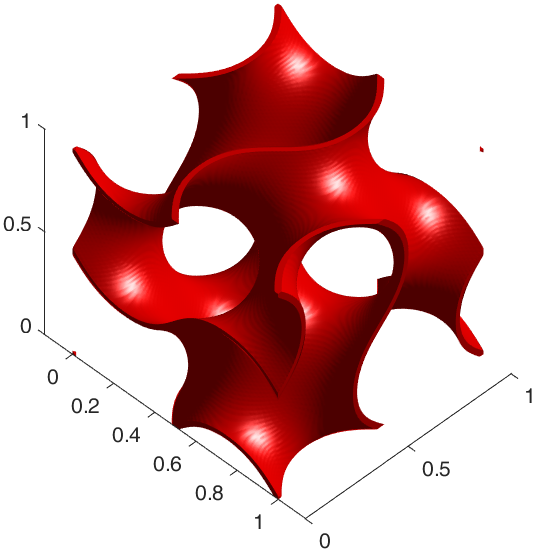}
        \caption{2\%}
    \end{subfigure}
    \begin{subfigure}[b]{0.25\linewidth}
        \includegraphics[width=\linewidth]{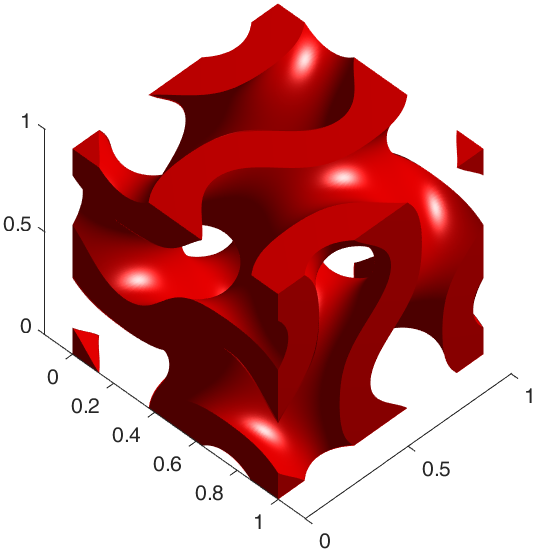}
        \caption{15\%}
    \end{subfigure}
    \begin{subfigure}[b]{0.25\linewidth}
        \includegraphics[width=\linewidth]{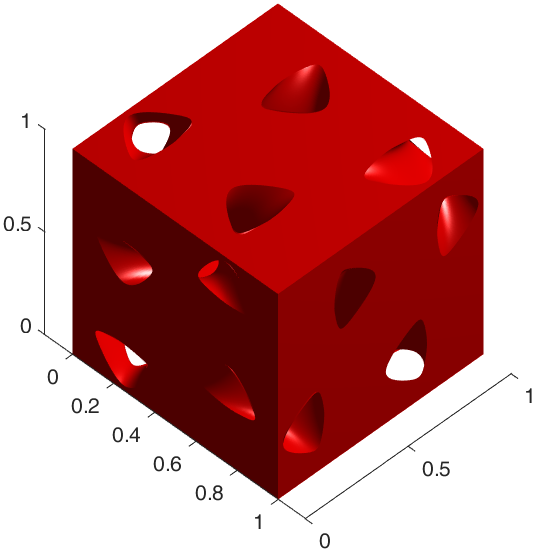}
        \caption{30\%}
    \end{subfigure}
    \caption{Sheet-network gyroid unit cell with varying wall thicknesses expressed as a percentage of unit cell size}
    \label{fig:gyroid-unit-cell}
\end{figure}

For the design region in the implant, we select a gyroid lattice structure. A gyroid is a form of TPMS that is usually approximated by the implicit equation
\begin{equation} \label{eq:gyroid}
    \begin{aligned}
          f(\mathbf{x}) = &\sin(x_1)\cos(x_2) + \sin(x_2)\cos(x_3) \\
          & + \sin(x_3)\cos(x_1) = 0,
    \end{aligned}
\end{equation}
where $x_1$, $x_2$, $x_3$ are spatial coordinates. The exact definition of the surface is complex (cf.~\cite{Schoen1970}), and the generated surface differs only slightly from that defined by \eqref{eq:gyroid}. An example of the approximated gyroid unit cell is presented in \reffig{fig:gyroid-surface}. The foregoing expression defines the base surface of the lattice; the solid lattice is the region defined by 
\begin{equation}
    \label{eq:thick-surface}
    |f(\mathbf{x})| \leq \delta.    
\end{equation}
It should be noticed, however, that the level-set value $\delta$ is approximately equal to the surface thickness $\tau$ for small values of $\delta$, but it differs from it otherwise. The solid given by \eqref{eq:thick-surface} is called a \textit{sheet-network} lattice, and it should be noted that it does not have uniform thickness. For simplicity, however, in this work we assume the surface thickness is uniform. The \textit{solid-network} TPMS lattice (which is not considered in this work) consists of entirely having solid material on one side of the level-set surface, i.e., $f(\mathbf{x}) \leq \delta$. Examples of sheet-network gyroid unit cells are shown in  \reffig{fig:gyroid-unit-cell} for various thickness values. 

TPMS lattices offer several advantages for scaffolds for tissue repair: they can be additively manufactured without internal support material thanks to the minimal surface property; they provide continuous networks of void space, allowing blood vessels and osteoblasts to infiltrate the lattice; they have a high specific surface area, providing ample space for bone deposition; and they are free from sharp corners and edges and thus from stress concentrations, reducing the required material for the load-bearing implant \cite{Lv2022}. 

The evaluation of the bone growth model requires a FEA of the bone-implant system at each time step, and thus requires the creation of a suitable mesh. This is problematic because the dimensions of the TPMS lattice features are very small relative to the dimensions of the entire bone-implant system, thus an impractically large number of elements would be required for a body-fitted mesh. This would greatly increase the computational cost of the analyses themselves, and would require a new mesh to be generated upon design changes. To circumvent this, and as is common in multi-scale design approaches (see \cite{wu2021}), the lattice is substituted with a homogenized material with equivalent properties. Consequently, the design region is modeled using a coarser mesh of solid elements that does not conform to the lattice small features. 

The properties of the equivalent material are obtained via numerical homogenization, which approximates the large-scale behavior of materials with periodic microstructures based on the response of the microstructure's unit cell to prescribed loads under periodic boundary conditions \cite{Liu2021, Hollister1992}. This technique effectively averages the behavior of the microstructure, but the assumption of periodicity means the results may not be valid near disruptions in the bulk lattice, such as at its the boundaries. For this work, we assume the homogenized properties are a reasonable approximation of the effective properties at any point in the design region. A surrogate model of the homogenized effective properties as a function of the TPMS surface thickness and the amount of bone ingrowth in the labyrinthic region of the lattice is subsequently constructed. For the FEA of the implant, the evaluation of homogenized material properties at a given point (e.g., the element centroid or an integration point) is thus performed by passing the thickness and bone density at that point to the surrogate model.

The numerical homogenization analysis for each combination of surface thickness and bone density is performed using a Matlab code adapted from the geometry projection method of \cite{kazemi2018}. In the geometry projection method, a high-level parametric description of the solid geometry (such as the one given by \eqref{eq:thick-surface}) is mapped onto a continuous pseudo-density field. The pseudo-density attains a value of 1 or 0 for elements fully inside or fully outside the solid surface, respectively, and an intermediate value at points on or near the boundary.This pseudo-density is subsequently used to interpolate the properties between the internal and external materials, in a manner similar to the solid isotropic material penalization (SIMP) employed in density-based topology optimization \cite{Bendsøe2004,NORATO2015}. This allows the use of a structured grid that is not body-fitted to perform the analysis, and thus precludes the need to recreate the mesh for different surface thicknesses; and it allows for the analysis of unit cells with a relatively high thickness, which would not be possible by meshing the surface with shell elements.

As in most geometry projection techniques, we assume for simplicity that the projected pseudo-density is uniform within each element in the mesh. Here, we do not compute the element pseudo-density based on the solid surface implicit representation of \eqref{eq:thick-surface}; instead, we compute it directly in terms of the signed distance $\phi_e$ from the centroid of element $e$ to the TPMS surface as 
\begin{equation}
\label{eq:geom_proj}
\rho_e\left(\phi_e \right) :=
\begin{cases}
0, & \text{if } \phi_e/r < -1 \\
\widetilde{H}(\phi_e/r), & \text{if } -1 \leq \phi_e/r \leq 1 \\
1, & \text{if }  \phi_e/r > 1, 
\end{cases}
\end{equation}
where $\mathbf{x}_e$ is the element's centroid, $\mathbf{z}$ is the vector of thickness design variables detailed in \refsec{subsec:design-rep}, $r$ is the radius of a fixed sampling window that contains the element, and $\widetilde{H}$ is a differentiable approximation of the Heaviside function given by
\begin{equation}
    \label{eq:smooth-Heaviside}
    \widetilde{H}(y; \eta) = \frac{1}{2} +\frac{y}{2 \eta} + \frac{1}{2\pi} \sin\left(\frac{\pi y}{\eta} \right),
\end{equation}
with $\eta$ a fixed parameter that determines the transition width. The signed distance $\phi_e$ is computed as a function of the distance $D_e$ to the TPMS base surface as
\begin{equation}
    \label{eq:signed-dist}
    \phi_e(\mathbf{x}_e) = \tau - D_e(\mathbf{x}_e),
\end{equation}
where $\tau$ is the (uniform) thickness of the unit cell, and $D_e$ is obtained from the numerical solution of the problem $d(\mathbf{x}_e) = \min_{\mathbf{c}_e} \| \mathbf{x}_e - \mathbf{c}_e \|$ subject to $f(\mathbf{c}_e) = 0$, where $\mathbf{c}_e$ is the point on the base surface closest to $\mathbf{x}_e$.

The pseudo-density of \eqref{eq:geom_proj} is subsequently used to interpolate material properties. In this work, the low pseudo-density regions are not treated as void, but rather assumed to be filled with bone at some density. The stiffness of the bone can thus be included in the homogenized properties, allowing the surrogate to also account for the effects of bone ingrowth. Here we employ the SIMP-like interpolation to compute the elasticity tensor for element $e$ as
\begin{equation}
    \label{eq:SIMP}
    \mathbf{C}_e(\rho_e) = \rho_e^p \mathbf{C}_{b} + (1-\rho_e^p) \mathbf{C}_{0}(\rho_{b}),
\end{equation}
where $\mathbf{C}_{0}$ and $\mathbf{C}_{b}$ are the elasticity tensors for the implant material and bone, respectively, and the power $p$ penalizes the effective stiffness so that it satisfies theoretical bounds for porous materials. The physical density of the ingrown bone is denoted by $\rho_{b}$, and it is assumed to uniformly fill the void region of the unit cell. 

\begin{figure} 
    \centering
    \includegraphics[width=0.85\linewidth]{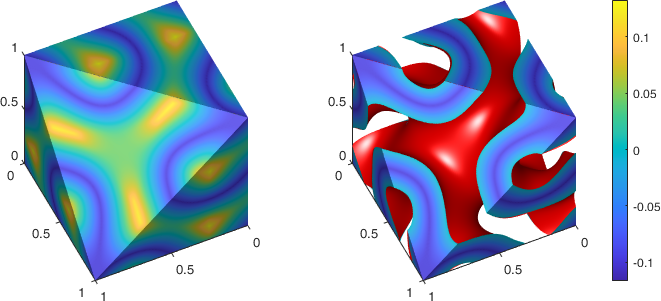}
    \caption{Gyroid signed distance field with a wall thickness of 20\%. The left image shows the field projected on the full unit cell volume, the right only on the interior of the solid portion. A corner is removed from both to show the interior}
    \label{fig:gyroid-sdf}
\end{figure}

To perform the homogenization analysis for a unit cell given by $(\tau, \rho_b)$, the surface thickness is first employed to compute the signed distance for each element using \eqref{eq:signed-dist}, which will determine which regions of the unit cell are solid and which are interstitial regions filled with bone, as shown in \reffig{fig:gyroid-sdf}.  
The zero value of the signed distance field defines the outer surfaces of the gyroid. To determine the specific surface area of the unit cell $S(\bfx)$, these surfaces were triangulated via Matlab's \texttt{isosurface} and their areas integrated to get the approximate total internal surface area. This value was then normalized by the unit cell volume to yield the specific surface area. The element signed distance $\phi_e$ is subsequently used to compute the element pseudo-density $\rho_e$ of \eqref{eq:geom_proj}; with this pseudo-density and the given bone density in the interstitial region, the elasticity tensor for each element in the mesh is computed using \eqref{eq:SIMP}. In addition to computing the effective elastic moduli of the homogenized unit cell, the pseudo-density field is also integrated over the unit cell volume to approximate the porosity of the unit cell when no bone is present in the interstitial space.

To generate the data necessary to build the surrogate, the homogenization analysis is performed on a 20-level full-factorial sampling of the lattice design space, with two factors: the surface thickness $\tau$ and the bone density $\rho_b$ in the interstitial region (for a total of 400 samples). A set of unit cells with wall thickness values ranging from 2\% to 40\% of the unit cell size and with ingrown bone densities ranging from 0.05 g/cc to 1.92 g/cc were numerically homogenized. These data were subsequently fit with a second-order spline interpolant provided by the SPLINTER library (\cite{SPLINTER}) to provide a continuously differentiable estimate of the unit cell properties.  The porosity and specific surface area values for the training simulations are similarly used to construct surrogates of the unit cell porosity and specific surface area as a function of surface thickness, $\xi(\tau)$ and $S(\tau)$ respectively. $\xi(\tau)$ is used to impose volume fraction constraints, as discussed in \refsec{subsec:optimization-problem}. $S(\tau)$ provides the local specific surface area required to compute $S_{d}(\bfx)$ from \eqref{eq:ssa-implant}. The value of $\tau$ at $\bfx$ is computed using the design representation, covered in \refsec{subsec:design-rep}.

\subsubsection{Minimum pore size}
\label{subsubsec:min-pore-size}

\begin{figure} 
    \centering
    \includegraphics[width=0.5\linewidth]{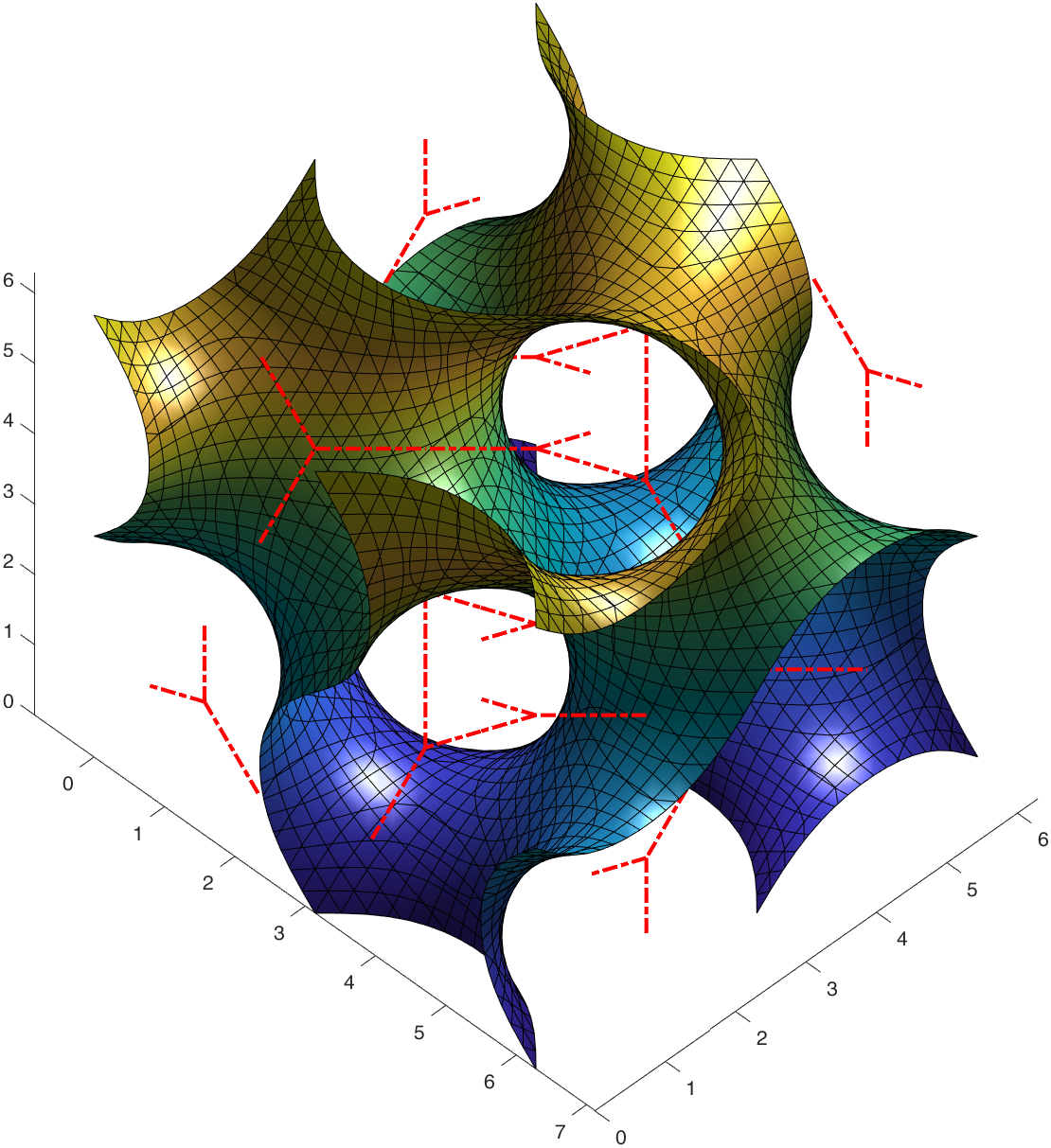}
    \caption{Gyroid base surface superimposed with dashed lines indicating the medial axis of the passages}
    \label{fig:gyroid-centerline}
\end{figure}

A key feature of the gyroid lattices that must be controlled is the minimum pore size. Determining the minimum pore size in a TPMS lattice is not straightforward, since the implicit representation of the surfaces given by \eqref{eq:thick-surface} does not readily allow to compute the radius of the largest sphere that passes through the lattice.  
There is no analytical solution for the minimum pore size that we are aware of. To determine what the minimum pore size would be, we thus employ a numerical approach. The distance values $D_e$ at the element centroids discussed in \refsec{subsec:gyroid-properties} are used to construct a surrogate of the distance $\tilde{d}(\mathbf{x})$ as a function of spatial coordinates. The $C^1$-continuous modified Akima interpolant (\cite{Akima1970, Ionita2019}) from MATLAB is used for this purpose.
This surrogate is subsequently used to perform a gradient-based optimization to find the point whose distance to the base surface is minimum and at which the norm of the gradient of the gyroid equation \eqref{eq:gyroid} is zero. The corresponding optimization problem is stated as
\begin{align} \label{eq:pore-size-opt}
     &p_s(\mathbf{x}) = \min_{\mathbf{x}} \tilde{d}(\mathbf{x})  \nonumber \\
     &\textrm{subject to}  \\
     &\|\nabla f(\mathbf{x})\| = 0. \nonumber 
\end{align}
The constraint on the gradient restricts the point $\mathbf{x}$ to lie on the medial axis of the void region (i.e., the gyroid passages), see \reffig{fig:gyroid-centerline}. That is, the optimal solution $\mathbf{x}^*$ is the point along the passage medial axis that is closest to any wall of the passage. Since the method searches for the minimum distance, and assuming the global optimum is found, every point other than $\mathbf{x}$ on the medial axis of the void region is at least as far from a wall. Therefore, the corresponding distance between the choke point $\mathbf{x}^*$ and the gyroid base surface determines the largest diameter sphere that could pass through the  labyrinth of the zero-thickness gyroid without jamming. We define the zero-thickness pore size $p_0$ as this diameter. The position of the choke point does not change with uniform-thickness extrusion of the base surface, hence this point can be used to determine the pore size $p_s$ for any surface thickness $\tau$ as $p_s(\tau) = p_0 - \tau$. 

\begin{figure}[!htbp]
    \centering
    \includegraphics[width=0.5\linewidth]{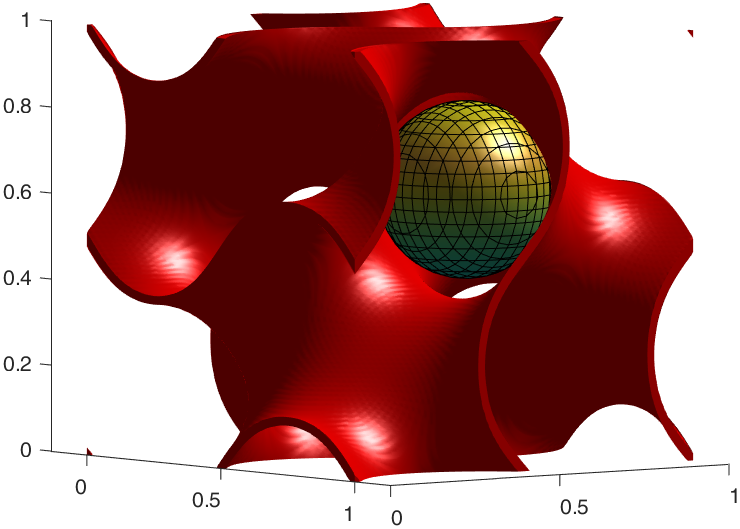}
    \caption{Maximum radius sphere that will pass through the gyroid labyrinth (2\% wall thickness)}
    \label{fig:gyroid-pore-size}
\end{figure}

Problem \eqref{eq:pore-size-opt} has multiple local minima, therefore the optimization is performed multiple times from different starting points to increase the chance of finding the global optimum. It should also be noted that, due to the symmetries of the gyroid unit cell, the choke point is not unique (i.e., the solution to problem \eqref{eq:pore-size-opt} only has weak global minima). The zero-thickness pore size $p_0$ was found to be ~43\% of the unit cell size. The size and location of the minimum pore size for a wall thickness of 2\% is shown for illustration in \reffig{fig:gyroid-pore-size}.

\subsection{Design Representation} 
\label{subsec:design-rep}

\begin{figure} 
    \centering
    \includegraphics[height=0.7\linewidth, angle=-90,origin=c]{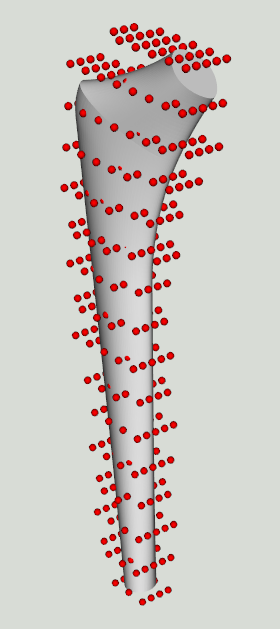}
    \vspace{-3cm}
    \caption{Thickness control points in red, superimposed on the gray design region. A total of 398 points are used}
    \label{fig:control-grid}
\end{figure}

To define the thickness field in the design region, a simple interpolation scheme is used. A spatial grid of control points is constructed that encompasses the entire design region, as shown in \reffig{fig:control-grid}. The thickness values at these points form the vector of design variables $\mathbf{z}$  of the optimization; at points that do not coincide with the grid control points, the thickness is determined via multidimensional linear interpolation.  Since the spacing of the control grid points is independent from the size of the elements in the finite element mesh, this approach precludes mesh-dependence of the optimal design. Moreover, the number of variables can be specified independently of mesh size, which helps control the cost of computing the sensitivities. This is particularly important as the sensitivities are herein calculated via the direct differentiation method, whose computational cost is proportional to the number of design variables. However, even when computing sensitivities using the adjoint method, in which the computational cost is less dependent on the number of design variables, assembling and tracking the sensitivity information still requires large amounts of cpu-time and storage \cite{Tortorelli1994}. Using fewer design variables is thus an effective means of reducing the required computational resources overall.  

Another advantage of the proposed thickness interpolation scheme is its compact support, meaning the thickness value at any point is only affected by a few of the nearest control points. Therefore, the sensitivity of the thickness at any point with respect to most of the design variables is zero for most problems. This sparsity can be exploited for considerable savings in both the computation and storage of the sensitivity information.

\subsection{Optimization problem} \label{subsec:optimization-problem}
The optimization problem we seek to solve is
\begin{equation}
    \begin{aligned}
        &\max_{\bm{z}} \quad m_f(\bm{z}) \\
        &\textrm{subject to} \\
        & V_{f}(\bm{z}) - V^* \leq 0 \\
        & \tau_{\min} \leq z_{i} \leq \tau_{\max} \quad \forall i \in 1..n\\
    \end{aligned}
\end{equation}
where $m_f(\bm{z})$ is the mass bone growth in the implant design region, and $\bm{z}$ is the vector of design variables consisting of the thicknesses at the $n$ control points, $z_1, \ldots, z_{n}$. The thickness values are bounded below by $\tau_{\min}$ to conform to the minimum thickness that can be manufactured, and above by $\tau_{\max}$, which ensures the minimum pore size is large enough to allow vascularization. For this work the minimum pore size is 100 $\mu$m based on \cite{Zhu2020}, while the minimum wall thickness is 300 $\mu$m. It is worth noting that the lower bound $\tau_{min}$ effectively imposes a lower bound on the stiffness of the implant, since the structural compliance strictly decreases with an increase in material. Even at this lower bound, the compliance of the implant is a fraction of the bone it replaces. This is at least partially due to the implant's volume being smaller than the replaced bone, as well as the much higher stiffness of titanium. 

A constraint is also imposed on the volume fraction $V_{f}(\bm{z})$, so that it is less than a specified upper bound $V^*$. $V_{f}(\bm{z})$ is defined as
\begin{equation}
    V_{f}(\bm{z}) = \frac{1}{V_{\Omega_{d}}}\int_{\Omega_{d}} \left( 1-\xi(\bm{z}) \right)  dV,
\end{equation}
where $\xi(\bm{z})$ is the unit cell porosity as defined in \refsec{subsec:gyroid-properties}, $\Omega_{d}$ is the implant region, and $V_{\Omega_{d}}$ is the total volume of that region. It should be noted that for a fixed unit cell size there are lower and upper bounds on the limit $V^*$ that can be imposed on the volume fraction, dictated by $\tau_{\min}$  and $p_s$ respectively.

\section{Implementation} \label{sec:implementation}
To demonstrate the proposed method, we optimize the femoral stem of a hip implant to maximize bone ingrowth. The software implementation for the bone growth simulation is the same as that of our previous work \cite{Cohen2024}, as are the partial derivatives of the bone growth with respect to the material properties. While a high level description is provided here, the interested reader is referred to that work for more details.

\subsection{Bone growth model}
\label{subsec:GeneralImplementation}
The finite element analysis to compute the strain energy density and subsequently the mechanical stimulus discussed in \refsec{subsec:bone-growth-model} is performed using an in-house code based upon the \texttt{deal.ii} finite element library \cite{dealII95}. Using the capabilities of the \texttt{deal.ii} library, we implemented a scalable distributed memory parallel finite element code. The \texttt{MUMPS} library is used via the \texttt{Amesos} package of \texttt{Trilinos} to solve the underlying systems of linear equations in parallel \cite{Trilinos2022, MUMPS}. The computation of the bone growth and its sensitivities are also parallelized. The sensitivities are computed via the direct method. For a detailed discussion of the sensitivity implementation, we refer the reader to \cite{Cohen2024}. A body-fitted first-order hexahedral mesh of the femur and implant system was created for the FEA using \texttt{Coreform Cubit} 2024.3 \cite{Cubit}. The mesh consists of approximately 40,000 elements. A mesh-size convergence study was performed, with the bone growth simulation first performed on the aforementioned mesh, and then on a uniformly refined mesh with approximately 320,000 elements. The difference in final mass bone growth between the two solutions was less 1\%, so the base mesh was deemed acceptable. The simulation is assumed to start ($t = 0$) immediately after implantation of the implant with no ingrown bone. To ensure the analysis problem is well-posed, and as outlined in \refsec{subsec:bone-growth-model}, a negligible but positive ersatz density of bone in the implant is assigned at $t = 0$. The duration of the simulation was 8 weeks, with a fixed time-step of 1 day.

The empirical constants required for the bone growth analysis in the case of humans are obtained from \cite{Carter2001}, and are listed in \reftable{tab:growth-constants}.
\begin{table*}[!htbp]
    \centering
	\begin{tabular}{c c c c c }
		\hline 
		Variable & Description & Values & Units & Source\tabularnewline
		\hline 
		$m$ & Empirical weighting factor for mechanical stimulus & 4 &  & \cite{Carter2001}\tabularnewline
		$\Psi^{*}$ & Reference mechanical stimulus & 50 & MPa / day & \cite{Carter2001}\tabularnewline
		$w$ & Dead zone half width of mechanical stimulus & 12.5 & MPa / day & \cite{Carter2001}\tabularnewline
		$c_{s}$ & Sensitivity to mechanical stimulus & 0.02 & $\mu$m/MPa & \cite{Jacobs1994}\tabularnewline
		$\hat{\rho}_{b}$ & Maximum achievable bone density & 1.92 & g/cc & \cite{Jacobs1994}\tabularnewline
		$\tilde{\rho}_{b}$ & Minimum allowed bone density & 0.05 & g/cc & \cite{Jacobs1994}\tabularnewline
		$S_{eff}^b$ & Effective specific surface area for a bone element & 0.2 &  & \cite{Sanz-Herrera2008}\tabularnewline
		$S_{eff}^s$ & Effective specific surface area of a implant element & 0.6 &  & \cite{Sanz-Herrera2008}\tabularnewline
        \hline
	\end{tabular}
	\caption{Bone growth model constants}
	\label{tab:growth-constants}
\end{table*}

\subsection{Femur and implant geometry}
 
\begin{figure}
    \centering
    \begin{subfigure}[b]{0.25\linewidth}
        \includegraphics[width=\linewidth]{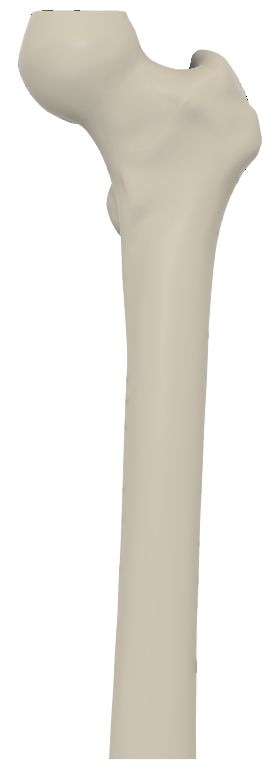}
        \caption{Intact femur}
        \label{fig:intact-bone}
    \end{subfigure}
    \begin{subfigure}[b]{0.25\linewidth}
        \includegraphics[width=\linewidth]{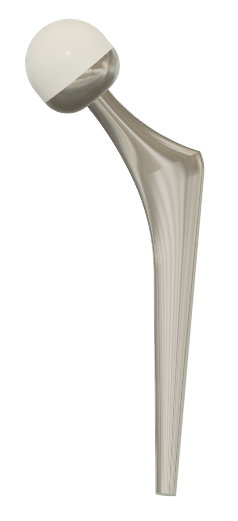}
        \caption{Implant}
        \label{fig:implant}
    \end{subfigure}
    \begin{subfigure}[b]{0.25\linewidth}
        \includegraphics[width=\linewidth]{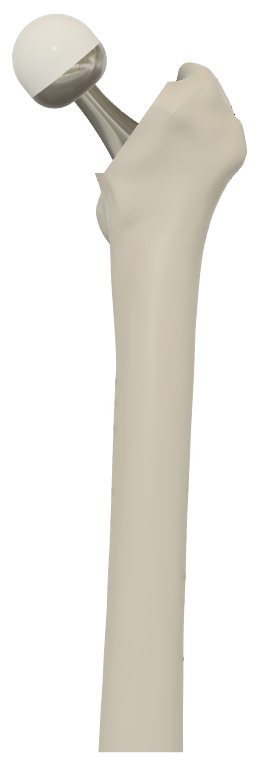}
        \caption{Assembly}
        \label{fig:assembly}
    \end{subfigure}
    
    \caption{Femur and implant geometry. Note that the white cap on the implant head (\reffig{fig:implant}, \reffig{fig:assembly}) is a low-friction plastic insert, and it is not included in the analysis. The top of the femur head on the intact bone (\reffig{fig:intact-bone}) is truncated by the limits of the CT scan. Since the femur head is removed from the analysis, its truncation has no effect on the results}
    \label{fig:Geometry}
\end{figure}

 The geometry from the human femur was derived from publicly available computed tomography (CT) scans provided by the National Library of Medicine's The Visible Human project \cite{VisibleHuman}. The CT scans are of a 72 year old male, and they have a voxel resolution 1 $mm^3$. The scanned femur was intact, with no implant (\reffig{fig:Geometry}). The CT scan was thresholded and segmented to isolate the bone of the femur, and a stereolithograpy (STL) file was generated using \texttt{Slicer} 5.6.1 \cite{Slicer561}. The CT scan was edited in this process to correct for an apparent movement of the patient during the scan, causing an abrupt discontinuity in the femur position. The STL was finally converted into a boundary representation (BREP) model and then modified to accept the implant using \texttt{Fusion 360}. The implant model itself is a generic computer-aided design (CAD) model of a cementless implant, which only consists of the femoral component.
 
\begin{figure}
     \centering
     \vspace{-1cm}
     \includegraphics[width=0.25\linewidth,angle=-90,origin=c]{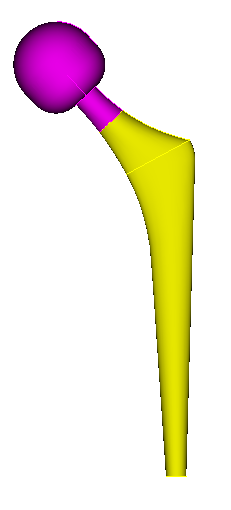}
     \vspace{-2.5cm}
     \caption{The yellow region of the implant is the design region $\Omega_{d}$, composed of a titanium gyroid lattice. The magenta region is the inert, non-designable region $\Omega_{i}$, composed of solid titanium. No bone growth is assumed to occur in $\Omega_{i}$}
     \label{fig:design-region}
 \end{figure}

 To obtain material properties for the bone, the bone density was estimated from the CT scan using the method outlined in \cite{Cohen2021}. This density was then used to estimate the properties of bone using empirical relationships, cf.~\cite{Carter2001, Cohen2024}. The bone is modeled as a solid isotropic material. The implant is assumed to be composed entirely of the titanium alloy Ti-6AL-4V (also isotropic), with an elastic modulus of 114 GPa and a Poisson's ratio of 0.3 \cite{He2018}. The implant is divided into a design and an inert region, shown in \reffig{fig:design-region}. The optimization only affects the geometry of the former region. Accordingly, the objectives and constraint values are computed based solely on this region. 
 
 The unit cell size is a constant for the optimization, set to 2.5 mm for this model. This strikes a compromise between the permissible thickness range set by the pore size and wall thickness constraints (cf. \refsec{subsec:optimization-problem}) and the overall size of the implant. Smaller unit cells necessarily have thinner walls and smaller pores, which reduce the usable range of thicknesses for a fixed minimum pore size and wall thickness. However, unit cells which are too large cease to be sufficiently periodic within the confines of the implant. This invalidates the assumption used to homogenize the TPMS properties, rendering them inaccurate. A unit cell size of 2.5 mm provides a usable thickness range from 12\% to 39\%  of the unit cell size, and ensures the the implant is spanned by at least 3 unit cells at its narrowest point.

\subsection{Boundary conditions}

For the bone growth simulation, boundary conditions are selected to best capture in-vivo loads. The force directions and magnitudes were computed based on the peak magnitudes and forces from \cite{Bergmann2016}, and the loading frequencies are obtained from \cite{Morlock2001}. Only one load case is used for the simulation, namely walking. While there are other situations where load magnitudes on the femur are higher, such as climbing stairs or standing up, the median number of loading cycles per day is typically less than 10 \% that of walking. Since the measure of mechanical stimulus used is frequency-weighted, walking dominates the total mechanical stimulus. Moreover, the directions of the forces at maximum load for these activities are very similar, so they promote bone growth in similar regions. Experiments with the bone growth simulation found that while including the loads corresponding to standing up and stair climbing greatly increased the simulation run-time, the amount of bone growth changed by only 2\%. 

\begin{figure}
    \centering
    \includegraphics[width=0.3\linewidth]{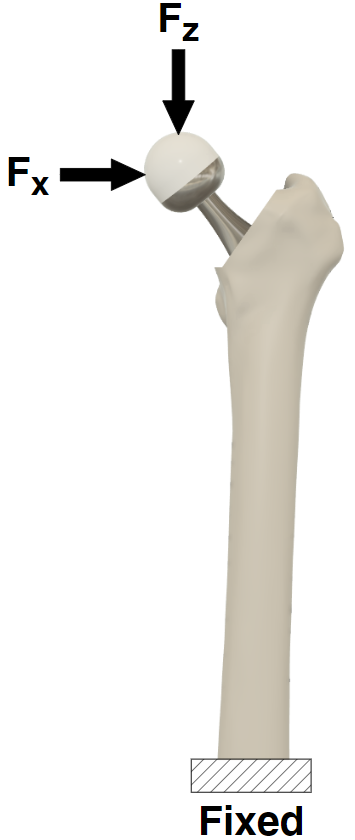}
    \caption{Boundary conditions. Note that a force $F_y$ (not shown in the figure) is also applied to the implant head pointing out of the page}
    \label{fig:boundary-conditions}
\end{figure}

All loads are applied to the head of the implant. The distal portion of the femur, where the bone joins the knee, is cut off (\reffig{fig:boundary-conditions}). An encastre boundary condition is placed on the cut face to anchor the bone femur assembly in place. The implant and femur are meshed using a single continuous mesh, which effectively corresponds to a bonded boundary condition. In practice, the implant is only press fit in place on implantation, and it may not osteointegrate over its whole surface. Therefore the system could be more accurately modeled as sliding contact with friction at the interface. However this was omitted to reduce the computational cost of the model and to retain the much simpler sensitivity analysis for the linear problem. Accordingly, stresses resulting from the press-fit itself are not considered. 

\section{Results} 
\label{sec:results}

\begin{figure}

    \centering
    \begin{tabular}{c c}
    \includegraphics[height=0.35\linewidth]{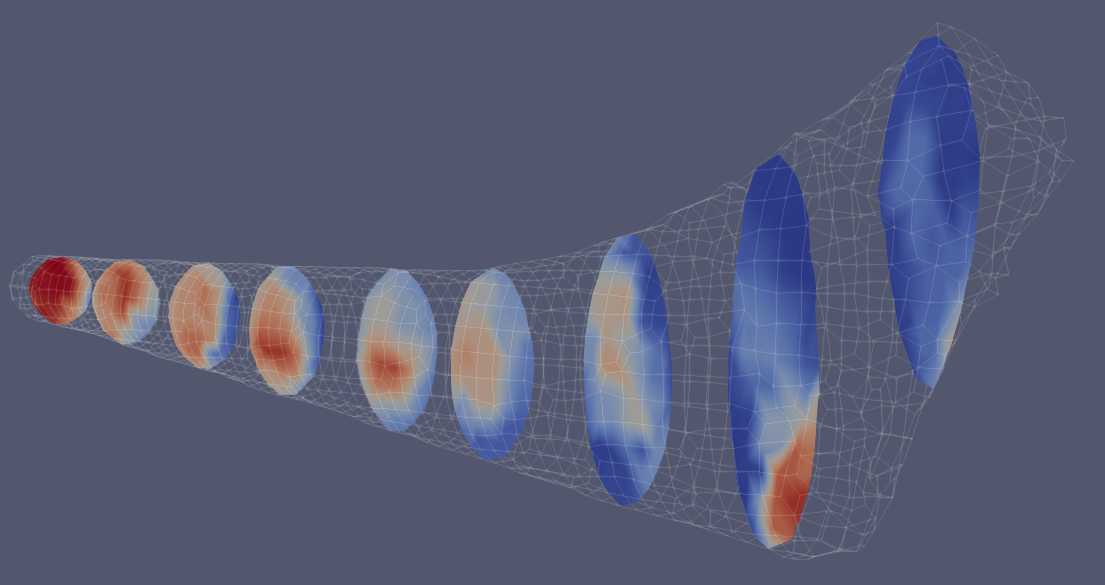} & 
    \includegraphics[height=0.35\linewidth]{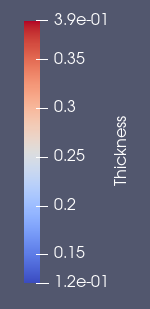}
    \end{tabular}
    \caption{Cross sections of the thickness field for $V^*$=70\%}
    \label{fig:CrossSections}
\end{figure}

\begin{figure}
    \vspace{-0.5cm}
    \centering
    \begin{tabular}{c c}
    \includegraphics[height=0.35\linewidth]{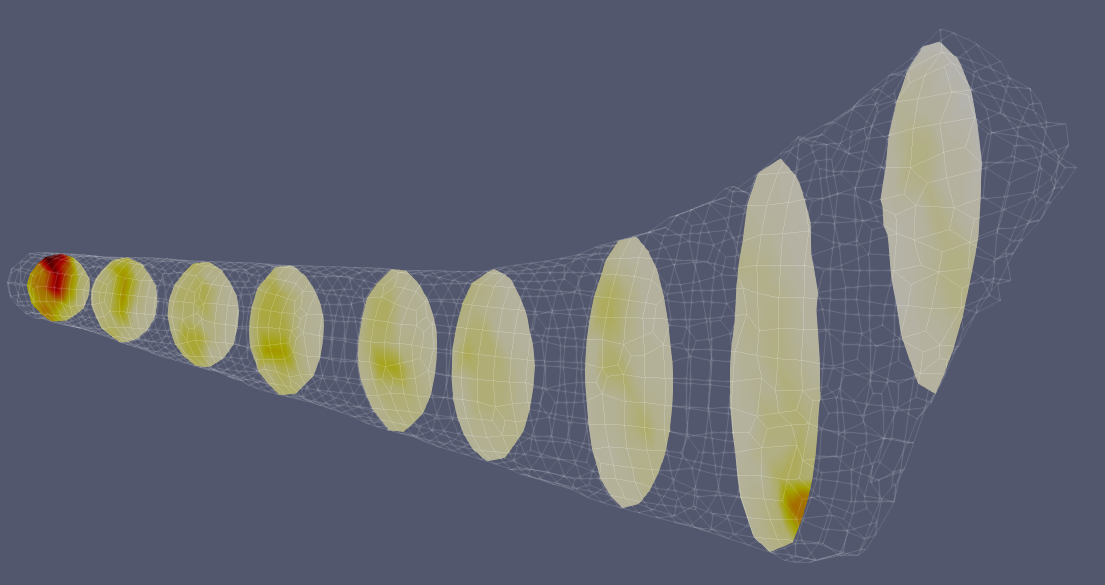} & 
    \includegraphics[height=0.35\linewidth]{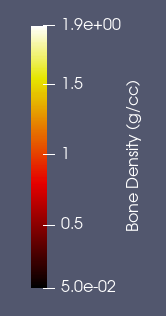}
    \end{tabular}
    \caption{Cross sections of the bone density field for $V^*$=70\%}
    \label{fig:BGCrossSections}
\end{figure}

We optimized the femur implant subject to volume fraction constraints of $V^*$ of 50\%, 60\%, and 70\%. For comparison purposes, we also performed bone growth simulations for the uniform-thickness designs of the same volume fractions, and for the uniform-thickness designs at the lower ($\tau_{\min}$) and upper ($\tau_{\max}$) bounds on the thickness, which correspond to the maximum and minimum allowable volume fractions, respectively. The optimal thickness and final bone density fields for the optimized designs are shown in \reftable{tab:OptimizationResults}, along with the corresponding fields for the thinnest- and thickest-allowable designs. \reffig{fig:CrossSections} and \reffig{fig:BGCrossSections} show the thickness and bone density fields respectively at several cross sections throughout the implant for the best performing design. The numerical results for mass growth for all the designs are presented in \reftable{tab:OptimizationNumbers}. A solid model of an optimized implant design with the gyroid microstructure is shown in \reffig{fig:ColoredSTL}.

\begin{table*}
    \setlength{\tabcolsep}{2pt}
    \newcommand{\imgheight}{1.1in}
    \newcommand{\colwidth}{0.15\linewidth}
    \centering
    \begin{tabular}{c | c c c c c}
        $V^* (\%)$ & Min allowable (36) & 50 & 60 & 70 & Max allowable (95)\\
        \hline
        {} & {} & {} & {} \\
        \savecellbox{$\tau$} & 
        \savecellbox{\begin{subtable}{\colwidth}
            \centering
            \includegraphics[height=\imgheight]{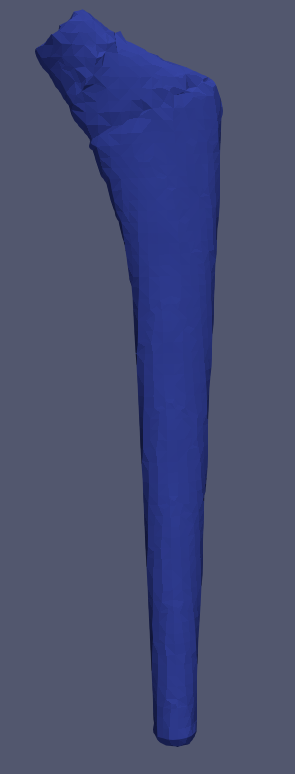}
            \includegraphics[height=\imgheight]{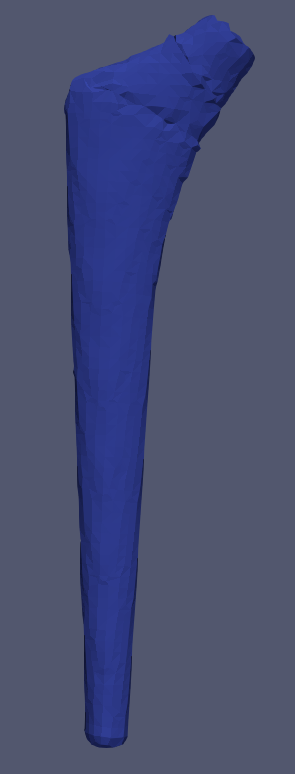}
            \caption{}
            \label{fig:VFMin}
        \end{subtable}}
        &
        \savecellbox{\begin{subtable}{\colwidth}
            \centering
            \includegraphics[height=\imgheight]{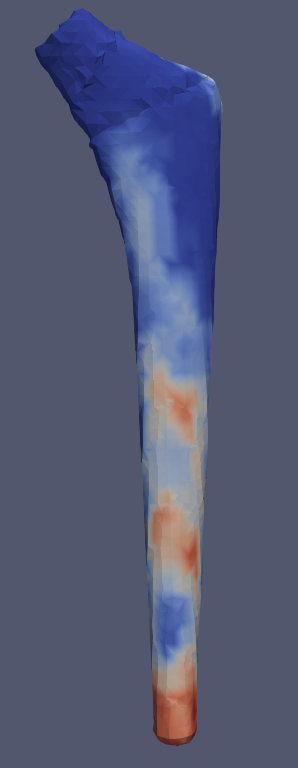}
            \includegraphics[height=\imgheight]{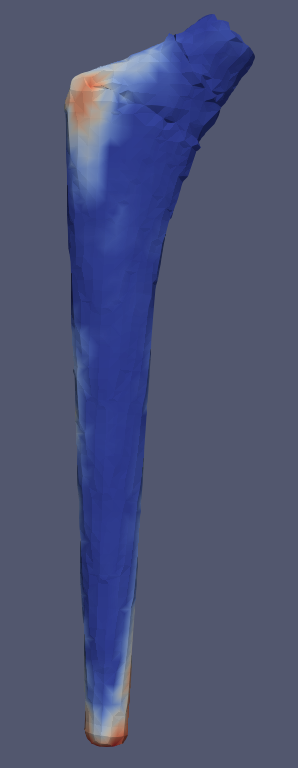}
            \caption{}
            \label{fig:VF50}
        \end{subtable}}
        &
        \savecellbox{\begin{subtable}{\colwidth}
                \centering
                \includegraphics[height=\imgheight]{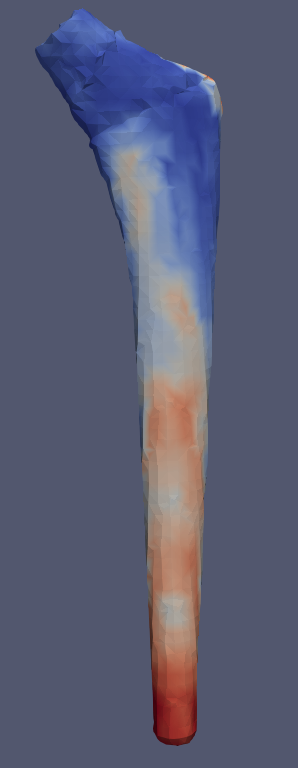}
                \includegraphics[height=\imgheight]{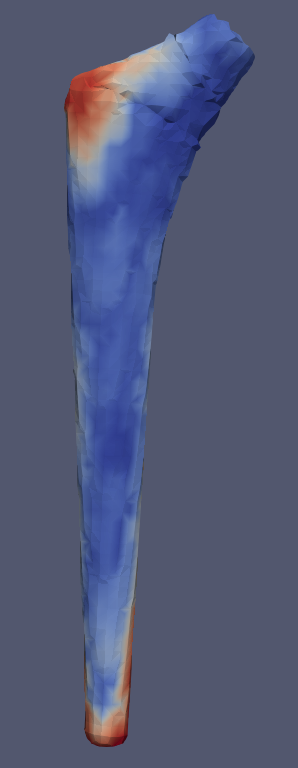}
                \caption{}
            \label{fig:VF60}
        \end{subtable}} 
        &
        \savecellbox{\begin{subtable}{\colwidth}
                \centering
                \includegraphics[height=\imgheight]{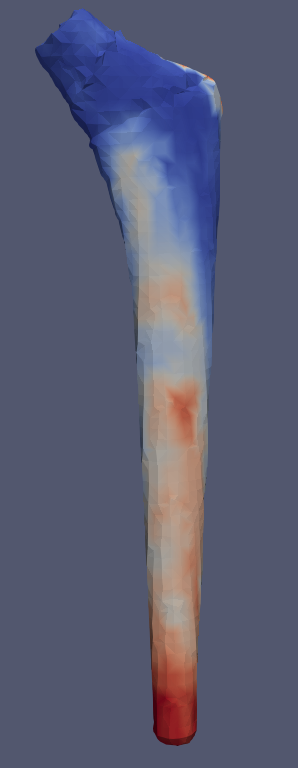}
                \includegraphics[height=\imgheight]{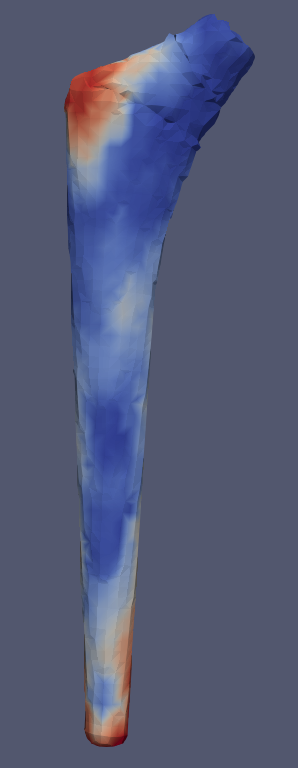}
                \caption{}
            \label{fig:VF70}
        \end{subtable}}
        &
        \savecellbox{\begin{subtable}{\colwidth}
            \centering
            \includegraphics[height=\imgheight]{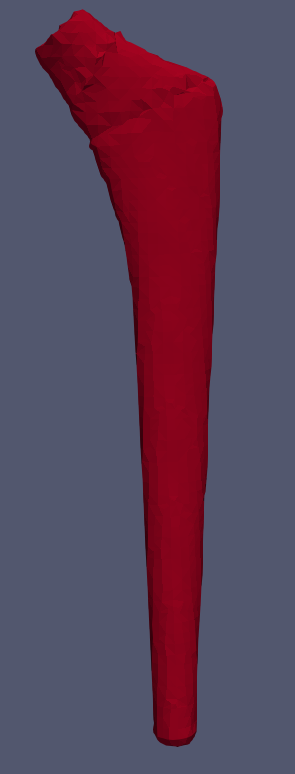}
            \includegraphics[height=\imgheight]{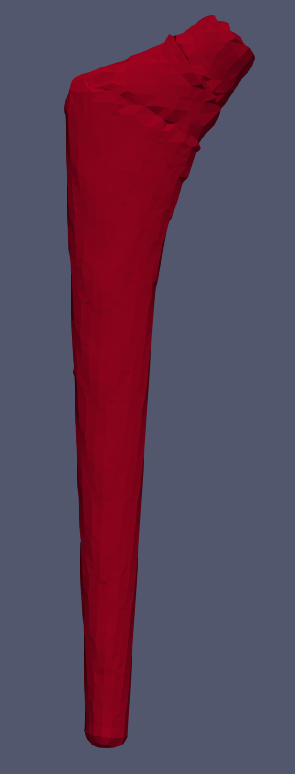}
            \caption{}
            \label{fig:VFMax}
        \end{subtable}}
        \\ [-\rowheight]
        \printcellmiddle &
        \printcellmiddle &
        \printcellmiddle &
        \printcellmiddle &
        \printcellmiddle &
        \printcellmiddle
        \\
        {} & \multicolumn{5}{c}{\includegraphics[width=0.4\linewidth]{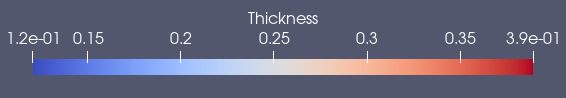}}
        \\
        \savecellbox{$\rho_b$} & 
        \savecellbox{\begin{subtable}{\colwidth}
            \centering
            \includegraphics[height=\imgheight]{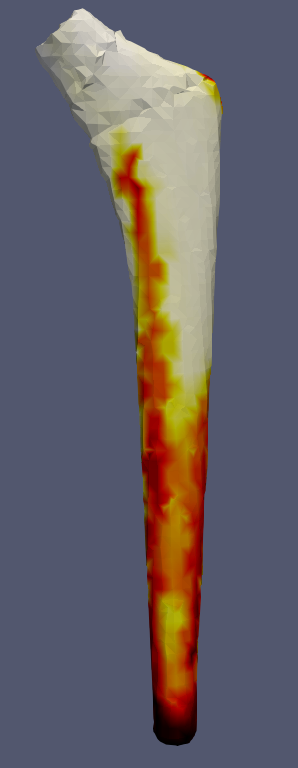}
            \includegraphics[height=\imgheight]{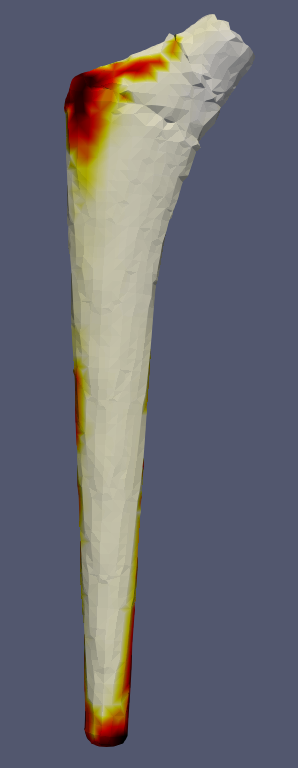}
            \caption{}
            \label{fig:VFMinBG}
        \end{subtable}}
        &
        \savecellbox{\begin{subtable}{\colwidth}
            \centering
            \includegraphics[height=\imgheight]{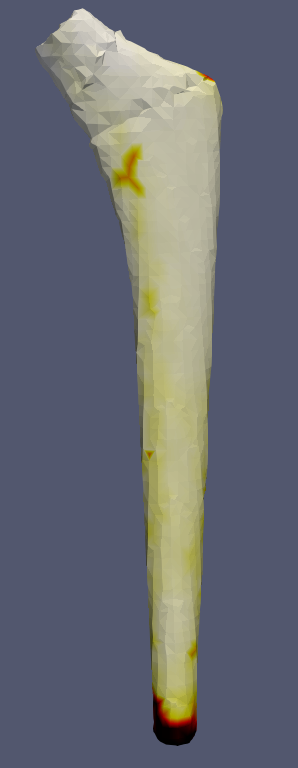}
            \includegraphics[height=\imgheight]{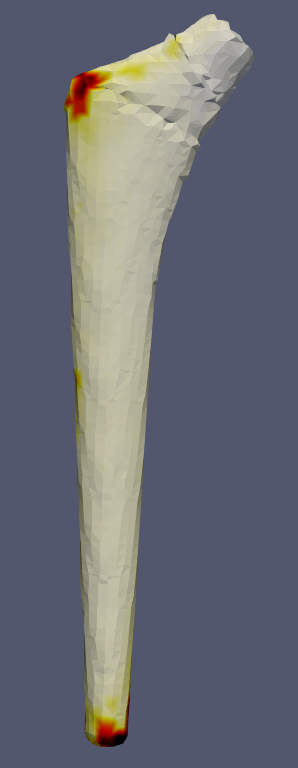}
            \caption{}
            \label{fig:VF50BG}
        \end{subtable}}
        &
        \savecellbox{\begin{subtable}{\colwidth}
                \centering
                \includegraphics[height=\imgheight]{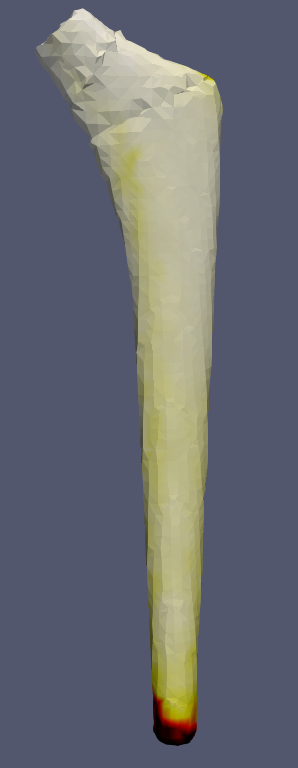}
                \includegraphics[height=\imgheight]{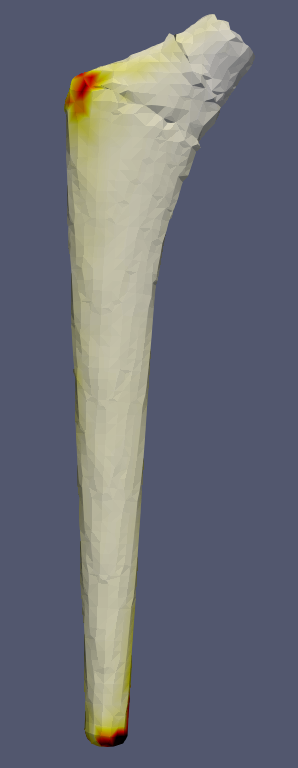}
                \caption{}
            \label{fig:VF60BG}
        \end{subtable}} 
        &
        \savecellbox{\begin{subtable}{\colwidth}
                \centering
                \includegraphics[height=\imgheight]{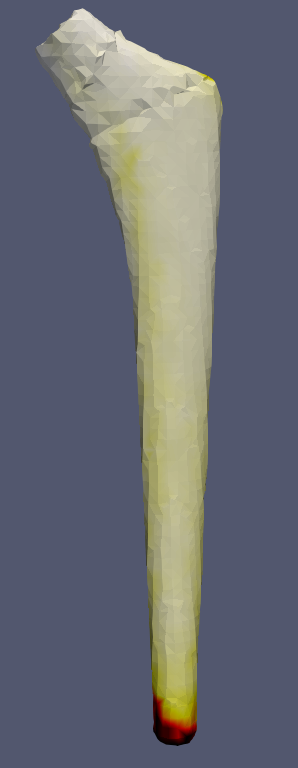}
                \includegraphics[height=\imgheight]{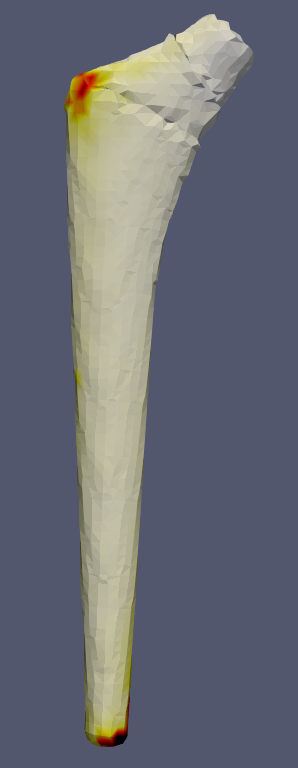}
                \caption{}
            \label{fig:VF70BG}
        \end{subtable}}
        &
        \savecellbox{\begin{subtable}{\colwidth}
            \centering
            \includegraphics[height=\imgheight]{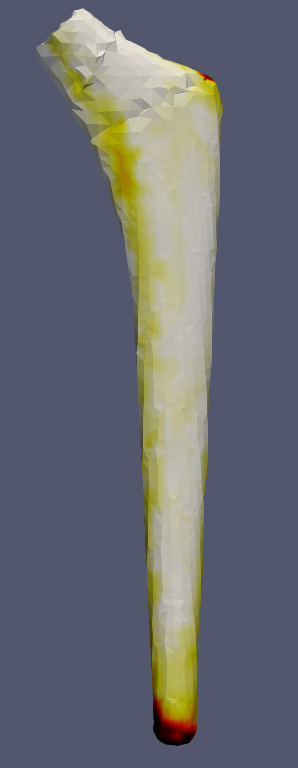}
            \includegraphics[height=\imgheight]{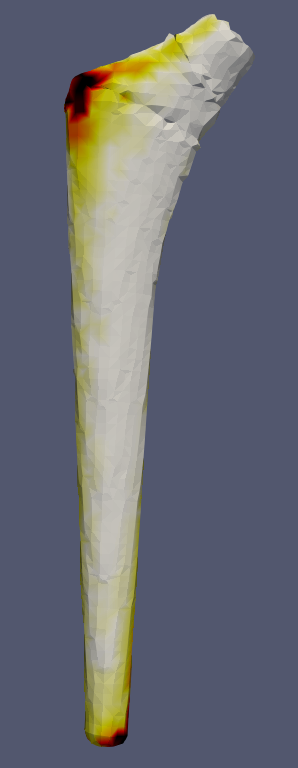}
            \caption{}
            \label{fig:VFMaxBG}
        \end{subtable}}
        \\ [-\rowheight]
        \printcellmiddle &
        \printcellmiddle &
        \printcellmiddle &
        \printcellmiddle &
        \printcellmiddle &
        \printcellmiddle
        \\
        {} & \multicolumn{5}{c}{\includegraphics[width=0.4\linewidth]{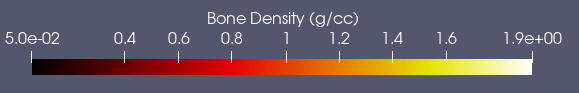}}
    \end{tabular}
    \caption{Front and back views of the optimization results for each volume fraction limit $V^*$. All optimizations are performed for 100 iterations. The top row ($\tau$) shows the thickness distribution as a percent of unit cell size, and the bottom row ($\rho_b$) shows the final bone density distribution in g/cc. For comparison, results for implants with the minimum and maximum allowable volume fraction (corresponding to the thickness bounds $\tau_{\min}$ and $\tau_{\max}$, respectively) are also shown}
    \label{tab:OptimizationResults}
\end{table*}

\begin{table*}
    \centering
    \setlength{\tabcolsep}{2.5pt}
    \small
    \begin{tabular}{c | c | c | c | c | c }
        $V^*$ (\%) & min & 50 & 60 & 70 & max\\
        \hline
        \begin{tabular}{c}
                $\tau$-field \\
                $\tau (\mu\textrm{m})$ \\
              $V_f$ (\%) \\
              $g_m$ (\%) \\
        \end{tabular}
        &
        \begin{tabular}{c}
            uniform \\
              300 ($\tau_{\min}$) \\
              36 \\
               78.0 \\
        \end{tabular}
        &
        \begin{tabular}{c c}
            uniform & optimized \\
              423.5 & - \\
              50 & 50 \\
               84.9 & 95.1 \\
        \end{tabular}
        &
        \begin{tabular}{c c}
            uniform & optimized \\
              519 & - \\
              60 & 60 \\
              89.0 & 96.1 \\
        \end{tabular}
        &
        \begin{tabular}{c c}
            uniform & optimized \\
              623.5 & - \\
              70 & 62 \\
             91.8 & 96.1 \\
        \end{tabular}
        &
        \begin{tabular}{c}
            uniform \\
              975 ($\tau_{\max}$) \\
              95 \\
            91.8 \\
        \end{tabular}
    \end{tabular}
    \normalsize
    \caption{Simulation results for each volume fraction limit $V^*$. $V_f$ is the volume fraction attained in the optimal design, and $g_m$ is the corresponding mass bone growth. The simulation results for the uniform-thickness designs are also shown for comparison.}
    \label{tab:OptimizationNumbers}
    
\end{table*}

{
\begin{figure}
    \newlength{\basewidth}
    \setlength{\basewidth}{345pt}
    \centering
    \begin{tikzpicture}
        \node[anchor=south west,inner sep=0] (detail) at (0,0) { \includegraphics[width=0.35\basewidth]{Images/Results/Thickness_Legend_Vertical.png}};
        
         \node[anchor=south west,inner sep=0] (detail) at (0,0.9\basewidth) { \includegraphics[width=0.35\basewidth]{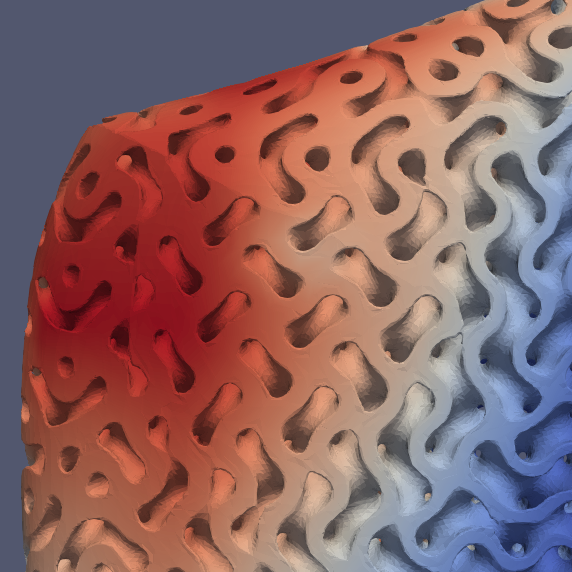}};
         
        \node[anchor=south west,inner sep=0] (main) at (0.4\basewidth,0) { \includegraphics[width=0.35\basewidth]{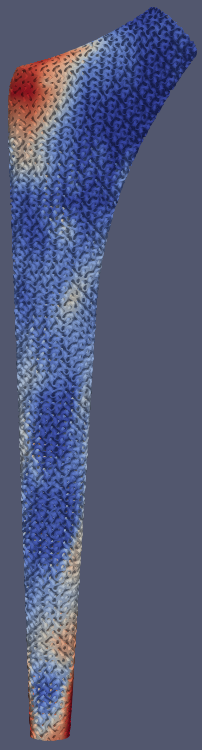}};
        
        \node (callout) at (0.47\basewidth, 1.15\basewidth) [draw,thick,minimum width=0.12\basewidth,minimum height=0.12\basewidth]{};
        
        \draw [black] (callout.north west) -- (detail.north east);
        \draw [black] (callout.south west) -- (detail.south east);
    \end{tikzpicture}
    \caption{Optimal design for $V^*=70\%$ colored by thickness (cf. ~\reftable{fig:VF70}) with gyroid microstructure rendered. The detail view on the left highlights the graded gyroid wall thickness. The gyroid geometry was generated using nTop \cite{NTop}}
    \label{fig:ColoredSTL}
\end{figure}
}

It is readily observed that all optimized thickness distributions are distinctly non-uniform, indicating that a spatially variable wall thickness is favorable over a uniform one. One way to confirm the benefit of a variable-thickness field is to compare the performance of the optimal designs to that of the implants having a uniform thickness equal to the lower and upper bounds $\tau_{\min}$ and $\tau_{\max}$, respectively. It can be observed from \reftable{tab:OptimizationResults} that the optimal designs outperform these uniform-thickness designs in terms of bone growth.

Another way to confirm the benefit of designing a variable-thickness lattice is to compare the results of the optimization for the 50\%, 60\%, and 70\% volume fractions to uniform-thickness designs that have the same volume fraction. The results of this comparison are shown in \reftable{tab:OptimizationNumbers}. Again, it can be observed that the optimal thickness-field designs outperform their uniform-thickness counterparts. It should be noted that since the volume fraction constraint for the $V^*=70\%$ design is inactive, the uniform-wall thickness design corresponding to $V^*=70\%$ has a larger mass.


The optimal designs all favor higher thicknesses near the bottom of the stem, and a decreased thickness towards the femoral head. This thickness grading increases the compliance towards the head of the implant, likely allowing more load to reach the lower portions of the stem and stimulate bone growth. All three designs have increased thickness around the outside corner of the stem. This seems to be an adaptation to draw more load to a relatively unstressed region. 

A similar dynamic seems to drive the generally higher thicknesses on the front of the implant compared to the back. Due to how the implant is placed inside the femur, the back side is closer to the stiff, high-density bone of the cortical shell. In a uniform thickness design, the majority of the load would therefore flow through the back face of the implant, leaving the front face comparatively under-stressed. Thinning out the back face while thickening the front face helps redirect some of the load, promoting bone growth in the front of implant. 

The optimized result for $V^*=70\%$ (\reftable{fig:VF70}) is particularly interesting since its volume fraction constraint is not active (it is 62\% for the optimal design). Unsurprisingly, its performance in terms of bone growth is negligibly better than that of the design with $V^*=60\%$. 
 It is likely the volume fraction constraint for the $V^*=70\%$ design is inactive due to a trade-off between unit cell compliance and specific surface area $S(\bm{x})$. For a fixed load, thickening the walls of sheet TPMS structures increases their stiffness and reduces the strain energy density which stimulates bone deposition. On the other hand, it also increases the amount internal surface area for bone deposition relative to the interstitial void space, allowing thicker TPMS unit cells to fill with bone more rapidly. Thus the optimal unit cell design for bone growth must balance these two factors, and the optimal design will depend on the magnitude of the load the unit cell sees. 

The bone density distributions (\reftable{tab:OptimizationResults}) and mass growth percentages (\reftable{tab:OptimizationNumbers}) of the optimized designs are all similar, with high density bone ($\geq$ 1.5 g/cc) throughout much of the implant stem. An example of a typical internal bone density distribution is shown in \reffig{fig:BGCrossSections}. The highest density bone is present on the outside of the implant, with lower densities on the implant interior. Both the maximum and minimum volume fraction designs (\reftable{fig:VFMaxBG}, \reftable{fig:VFMinBG}) have generally lower bone densities at the end of the simulation period, particularly the latter. 

Surprisingly, all of the designs show bone growth not just in regions enveloped by the femur, but into the exposed neck of the implant. It is unclear if bone deposition could occur so far from natural bone, since osteoblasts and blood vessels could only reach these regions by deeply infiltrating the tortuous gyroid passages. Bone deposition in the neck would likely be delayed because of this, if it occurred at all. The bone growth model does not currently account for this. The original version of the bone growth model by \cite{Sanz-Herrera2008} tried to model cellular infiltration/vascularization using a diffusion analysis, with the preexisting bone treated as infinite source of osteoblasts. However, this model had no sinks, i.e., the osteoblasts were never depleted by any mechanism. As a result, the implant (or scaffold in the original work) would quickly become saturated with cells, thus the diffusion analysis had little effect on the final results unless the structure's initial permeability was extremely low. While omitted from this work for that reason, a more sophisticated diffusion model with a depletion mechanism might provide useful insights.

The optimized designs offer clear benefits over their uniform thickness counterparts. The aforementioned greater bone growth would help ensure fixation of the implant; the optimized designs are also more compliant. The optimized designs' compliance exceeds their uniform thickness counterparts by 12\% for $V^*=50\%$, 18\% for $V^*=60\%$, and 26\% for $V^*=70\%$. This would likely help avoid stress shielding by the implant, and reduce resorption of the surrounding bone. 

\section{Conclusion} \label{sec:conclusion}

This work describes a method for the optimization of TPMS lattice structures for bone growth. A mechanobiological simulation of bone growth is used to directly evaluate the bone growth into the structure. This simulation takes into account the effects of the in-vivo loads, the frequency of those loads, the density distribution and geometry of the surrounding bone, and the bone growth into the structure over time. The proposed method tailors the structure to the patient by spatially varying the local unit cell wall thickness throughout the design region. The method was demonstrated on a cementless hip implant, with the goal of maximizing bone growth into the implant stem subject to constraints on volume fraction, the minimum wall thickness and minimum pore size. The optimizations with volume fraction limits of 50, 60 and 70\% produced variable-thickness designs with greater bone in-growth and compliance compared to uniform-thickness designs of the same volume fraction. This likely indicates the optimized designs would have better fixation and experience less resorption of the surrounding bone than the uniform thickness designs, thus reducing the risk revision surgery will be required.

Several major aspects of implant design remain to be considered. This method does not consider mechanical failure due to yielding, buckling or fatigue. All three of these factors must be incorporated to ensure the structural adequacy of the implant. Moreover, this method does not model bone resorption directly, another key factor of long term fixation. Such considerations are deferred to future work.

\section{Replication of Results}
Source code and data used for these results will be made available upon request.

\printbibliography

\end{document}